\documentclass[reprint,10pt]{revtex4-2}
\usepackage{braket}
\usepackage{mathtools}
\usepackage{adjustbox}
\usepackage[utf8]{inputenc}
\usepackage{graphicx}

\usepackage{csquotes}
\usepackage[dvipsnames]{xcolor}
\usepackage{color}
\usepackage{tikz}
\usepackage{graphics}
\usepackage{float}
\usetikzlibrary{shapes.geometric,backgrounds,positioning,shapes.geometric,decorations.markings,decorations.pathreplacing,arrows,knots,hobby,angles,quotes}
\usepackage{amsmath}
\usepackage{amssymb}
\usepackage{amsthm}
\usepackage{braket}
\usepackage{bbm}
\usepackage{siunitx}

\usepackage{subcaption}
\usepackage{caption} 

\DeclareMathOperator{\tr}{tr}
\DeclareMathOperator{\can}{\mathrm{CAN}}
\DeclareMathOperator{\rxx}{\mathrm{RXX}}
\DeclareMathOperator{\ryy}{\mathrm{RYY}}
\DeclareMathOperator{\rzz}{\mathrm{RZZ}}

\definecolor{dqnn1}{RGB}{120,94,240}
\definecolor{dqnn2}{RGB}{100,143,255}
\definecolor{qaoa1}{RGB}{254,97,0}
\definecolor{qaoa2}{RGB}{255,176,0}

\newcommand{\calibrationlineNRQNN}{\raisebox{2pt}{\tikz{\draw[-,dqnn1,dash dot,line width = 1.5pt](0,0) -- (5mm,0);}}}
\newcommand{\trainingCostMarkNRQNN}{\raisebox{-1pt}{\tikz{
    \node[mark size=3.5pt,color=dqnn1] at (2.5mm,0) {\pgfuseplotmark{triangle*}};
}}}
\newcommand{\validationCostMarkNRQNN}{\raisebox{-1pt}{\tikz{
    \node[mark size=3pt,color=dqnn1] at (2.5mm,0) {\pgfuseplotmark{*}};
}}}
\newcommand{\calibrationlineQAOA}{\raisebox{2pt}{\tikz{\draw[-,qaoa1,dash dot,line width = 1.5pt](0,0) -- (5mm,0);}}}
\newcommand{\trainingCostMarkQAOA}{\raisebox{-1pt}{\tikz{
    \node[mark size=3.5pt,color=qaoa1] at (2.5mm,0) {\pgfuseplotmark{triangle*}};
}}}
\newcommand{\validationCostMarkQAOA}{\raisebox{-1pt}{\tikz{
    \node[mark size=3pt,color=qaoa1] at (2.5mm,0) {\pgfuseplotmark{*}};
}}}

\newcommand{\calibrationline}{\raisebox{2pt}{\tikz{\draw[-,black,dash dot,line width = 1.5pt](0,0) -- (5mm,0);}}}
\newcommand{\trainingCostMarkOnly}{\raisebox{-1pt}{\tikz{
    \node[mark size=3pt,color=black] at (2.5mm,0) {\pgfuseplotmark{triangle*}};
}}}
\newcommand{\validationCostMarkOnly}{\raisebox{-1pt}{\tikz{
    \node[mark size=2.5pt,color=black] at (2.5mm,0) {\pgfuseplotmark{*}};
}}}
\newcommand{\trainingCostMark}{\raisebox{-1pt}{\tikz{
    \draw[-,black,line width = 1pt](0,0) -- (5mm,0);
    \node[mark size=3pt,color=black] at (2.5mm,0) {\pgfuseplotmark{triangle*}};
}}}
\newcommand{\validationCostMark}{\raisebox{-1pt}{\tikz{
    \draw[-,black,line width = 1pt](0,0) -- (5mm,0);
    \node[mark size=2.5pt,color=black] at (2.5mm,0) {\pgfuseplotmark{*}};
}}}
\newcommand{\calibrationCostMark}{\raisebox{-1.4pt}{\tikz{
    \fill[-,black, fill=black, line width = 1.5pt, opacity=0.1](0,0) -- (0, 2.5mm) -- (2.5mm, 2.5mm) -- (2.5mm, 0);
    \draw[-,white!50.1960784313725!black,line width = 1.5pt](0,2.5mm) -- (2.5mm, 2.5mm);
}}}

\tikzset{bordercolor/.style={
    draw={dqnn1},very thick,
    postaction = {draw,qaoa1,dash pattern= on \dashpattern off \dashpattern,dash phase=\dashphase,very thick}
}}

\newcommand{\dashpattern}{11.75pt}
\newcommand{\dashphase}{0pt}

\usepackage{hyperref}

\usepackage{cleveref}
\crefname{section}{section}{sections}
\Crefname{section}{Section}{Sections}
\crefname{equation}{Eq.}{Eq.}
\Crefname{equation}{Eq.}{Eq.}
\crefname{figure}{Fig.}{Fig.}
\Crefname{figure}{Fig.}{Fig.}
\crefname{appendix}{appendix}{appendices}
\Crefname{appendix}{Appendix}{Appendices}

\usepackage{pgfplots}
\pgfplotsset{compat=1.14}

\usepackage{amsfonts} 
\usepackage{lmodern,adjustbox}
\usepackage[skins,breakable]{tcolorbox}
\usepackage{tikz}
\usetikzlibrary{quantikz}
\tcbuselibrary{listings}
\tcbuselibrary{breakable}
\usepackage{changepage}

\usetikzlibrary{decorations.markings}

\hypersetup{pdfauthor={Kerstin Beer, Daniel List, Gabriel Müller, Tobias J. Osborne, and Christian Struckmann},pdftitle={Training Quantum Neural Networks on NISQ devices}}

\begin{document}

\title{Training Quantum Neural Networks on NISQ Devices}

\author{Kerstin Beer}
\author{Daniel List}
\author{Gabriel Müller}
\email{mueller-gabriel@outlook.de}

\author{Tobias J.\ Osborne}

\author{Christian Struckmann}
\email{christian.struckmann@gmx.de}

\affiliation{Institut f\"ur Theoretische Physik, Leibniz Universit\"at Hannover, Germany}
\maketitle

\textbf{
The advent of noisy intermediate-scale quantum (NISQ) devices offers crucial opportunities for the development of quantum algorithms. Here we evaluate the noise tolerance of two quantum neural network (QNN) architectures on IBM's NISQ devices, namely, \emph{dissipative QNN} (DQNN) whose building-block perceptron is a completely positive map, and the \emph{quantum approximate optimization algorithm} (QAOA). We compare these two approaches to learning an unknown unitary. While both networks succeed in this learning task, we find that a DQNN learns an unknown unitary more reliably than QAOA and is less susceptible to gate noise.
}

\section{Introduction}
%
The classical simulation of many body quantum systems is fundamentally limited by the exponential growth -- in the system size -- of the required computational resources. It was Feynman who first understood that what would become quantum computing can overcome this obstruction and offer classically unattainable computational possibilities \cite{feynman1982simulating}. This dream has come much closer to reality, particularly in the past year, as progress on engineering quantum computers has accelerated rapidly. In parallel, the exploration of various quantum algorithms is progressing steadily, already leading to early implementations of \emph{quantum machine learning} \cite{ciliberto2018quantum, dunjko2018machine, cerezo2020variational}.

Quantum machine learning models executed on quantum computers are expected to outperform their classical counterparts in numerous tasks (see, e.g., \cite{ciliberto2018quantum, kouda2005qubit} and references therein for a cross-section of results). Such models are largely implemented as parameterized quantum circuits \cite{bu2021statistical, benedetti2019parameterized, mitarai2018quantum, du2018expressive} which are executed on a quantum computer and whose parameters are  variationally optimized via classical side computation  \cite{stokes2020quantum, schuld2019evaluating, ostaszewski2019quantum, mitarai2018quantum}.

In recent years a variety of parametrized quantum circuit architectures for quantum machine learning have been proposed \cite{cerezo2020variational}.
Prominent amongst these are \emph{quantum neural networks} (QNNs), defined by quantum analogy with classical neural networks.
Their fundamental building block, the \emph{quantum perceptron}, has been defined in many ways \cite{beer2020training,schuld2020circuit, zhang2020toward, tacchino2020quantum, sharma2020trainability, skolik2020layerwise, torrontegui2019unitary, killoran2019continuous, steinbrecher2019quantum, farhi2018classification, cao2017quantum, wan2017quantum, da2016quantum, schuld2015simulating, altaisky2001quantum, lewenstein1994quantum}. QNNs have been successfully used to learn unitaries \cite{beer2020training, geller2021experimental}, perform classifications tasks \cite{zhang2020toward, schuld2020circuit, schuld2019quantum}, and denoise quantum data \cite{bondarenko2020quantum, achache2020denoising, wan2017quantum}, to name a few applications. The focus of this paper is on the model of \cite{beer2020training}, who define a quantum perceptron to be a completely positive map, leading to \emph{dissipative} QNN (DQNN) \footnote{Note that, contrary to what the name perhaps suggests, DQNN are universal for quantum computation and can equally simulate both unitary and dissipative quantum processes.}.
An alternative parametrized quantum circuit class which has attracted considerable interest for its practical benefits in implementations is that arising in the \emph{quantum approximate optimization algorithm} (QAOA), which features a sequence of alternating unitary operators \cite{farhi2014quantum, farhi2016quantum, hadfield2019quantum}. This variational class of quantum circuits has led to promising results for learning unitaries \cite{kiani2020learning} and solving combinatorial problems \cite{streif2019comparison,wang2018quantum,wecker2016training,maciejewski2021modeling,otterbach2017unsupervised,lechner2020quantum,farhi2015quantum,farhi2017quantum,lin2016performance}.

Successfully executing QNNs and the QAOA on today's \emph{noisy intermediate-scale quantum} (NISQ) devices \cite{preskill2018quantum} remains extremely challenging \cite{biamonte2017quantum}.
When exploiting gradient-based training methods, the presence of Barren plateaus, i.e., exponentially vanishing cost gradients, can significantly diminish their performance \cite{wang2020noise, cerezo2020cost, grant2019initialization, mcclean2018barren}. The increasing noise levels in higher-depth quantum circuits also poses a major challenge for accurately computing costs and gradients \cite{maciejewski2021modeling, alam2019analysis, xue2019effects, preskill2018quantum}. Thus, any approach to implementing these algorithms on currently available quantum hardware must be evaluated and optimized for their noise tolerance.

In this paper we exploit, and compare, DQNN and QAOA on current quantum hardware. The task we evaluate is that of learning an unknown unitary operator. The DQNN is implemented as a parameterized quantum circuit which is trained via gradient descent; the architecture of the resulting quantum circuit is optimized to mitigate the effect of noise. We compare the DQNN circuit with a similar implementation via the QAOA \cite{farhi2014quantum}, where the QAOA is described as a QNN. A notable difference between these approaches, is how the input information propagates through the network. In the case of DQNN, perceptron maps act on layers of different qubits, whereas the QAOA defines them as a sequence of operations on the same qubits. Both quantum circuits are implemented via Qiskit \cite{Qiskit} and executed on simulated and real quantum devices hosted by IBM \cite{ibmqe}.

Our paper is organized as follows:
In \Cref{chap2}, we review the network model and the corresponding quantum circuit for DQNN and the QAOA.
\Cref{chap3} features the architecture-specific details and implementation of learning via gradient descent.
In \Cref{chap4} we present our numerical results, where we compare both networks in terms of their robustness to noise and their generalization capabilities. Concluding remarks are found in \Cref{chap5}.

\begin{figure*}[ht]
    \centering
    \rule{\linewidth}{0pt}
    \includegraphics[width=\textwidth]{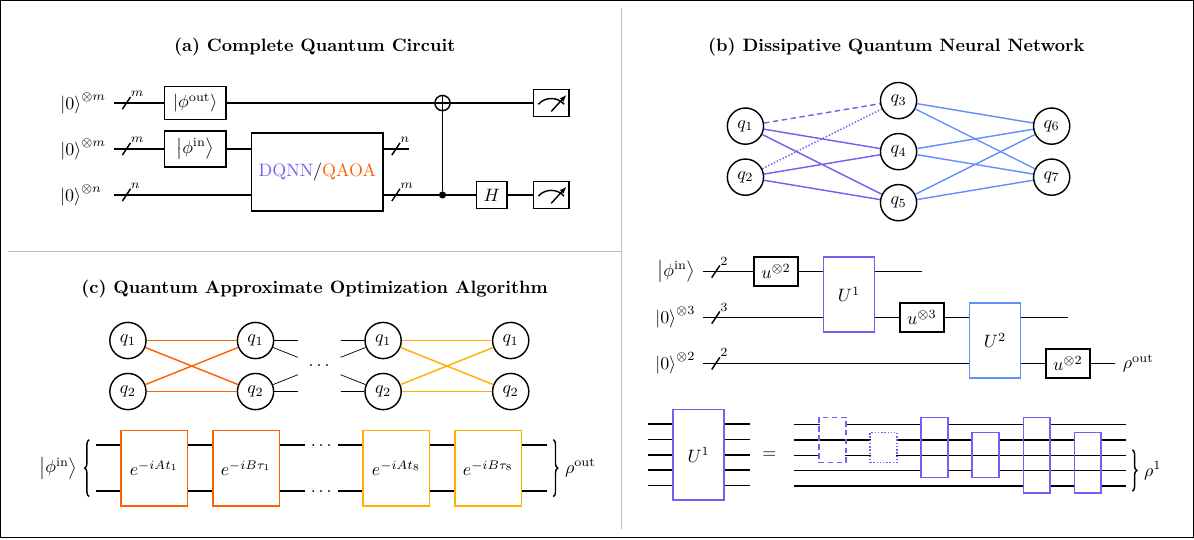}
    \caption{\textbf{Implementation of the DQNN and the QAOA.} The complete quantum circuit (a) consists of three parts: i) the initialization of the first two $m$ qubits in the target output state $\ket{\phi^\mathrm{out}}$ and input state $\ket{\phi^\mathrm{in}}$, respectively, ii) a QNN circuit that implements either the DQNN ($n \geq m$) or the QAOA ($n=0$), and iii) the fidelity measurement between the first $m$ and the last $m$ qubits via the destructive swap test. The DQNN (b) features one input layer, $L$ hidden layers, and one output layer with $m_l$ neurons, respectively. Each neuron is represented by an individual qubit. Two neighboured layers, $l-1,l$, are connected through the layer unitary $U^l$. The QAOA (c) consists of a sequence of alternating unitary operators $e^{-iAt_l}, e^{-iB\tau_l}$ with $l=1,\dots,N$. It can be interpreted as an $\left(N+1\right)$-layer QNN with $m$ neurons per layer where the operator pair $l$ connects the layers $l-1,l$. The quantum circuit only requires $m$ qubits such that all unitaries act on the same qubits.}
    \label{fig:circuits}
\end{figure*}
\section{Network Models}\label{chap2}
In this section we review DQNNs and the QAOA and, for each network, describe the network model and its implementation as a quantum circuit.

\subsection{Dissipative QNN}
The general structure of a DQNN is described in \cite{beer2020training}. It features one input layer ($l=0$), $L\geq 0$ hidden layers ($l=1,2, \dots, L$), and one output layer ($l=L+1$). The input and output layers consist of $m=m_0=m_{L+1}$ neurons, respectively, while the hidden layers consist of an arbitrary number of neurons $m_l \geq 1$.
(Generally, $m_{L+1} \neq m_0$ is possible but would correspond to a non-unitary transformation, which is out of the scope of the current paper.)
The input neurons are initialized in a given state $\ket{\phi^\mathrm{in}}$ while the neurons of the remaining layers are initialized in the vacuum product state $\ket{0...0}_l$. 
Two neighboured layers, $l-1$ and $l$, are fully connected via the layer unitary $U^l=U^l_{m_l}...U^l_1$. The quantum perceptron $U^l_j$ is defined neuron-wise as a general unitary operator acting on all $m_{l-1}$ neurons of layer $l-1$ and the $j$th neuron of layer $l$.
Layer by layer, the unitary $U^l$ is applied to all input states (layer $l-1$) and output states (layer $l$). 
The states related to layer $l-1$ are traced out afterward leaving only the output state of the remaing qubits as an input for the next layer. The resulting evolution is hence given, in general, by a completely positive map. This iterative process induces the feed-forward nature of the network. 
Thus, the network's final output state $\rho^{\mathrm{out}}$ is obtained by:
\begin{equation} \label{output_state_definition}
	\rho^{\mathrm{out}} \equiv \tr_\mathrm{in,hid} \left( \mathcal{U}\left(\rho^\mathrm{in} \otimes |0\cdots0\rangle_{\text{hid,out}}\langle 0\cdots 0|\right) \mathcal{U}^\dag\right),
\end{equation}
where the network unitary is $\mathcal{U}=U^{L+1}...U^1$ and the input state $\rho^\mathrm{in} = \ket{\phi^\mathrm{in}}\bra{\phi^\mathrm{in}}$.

\subsubsection*{The quantum circuit}
For our concrete implementation of a DQNN we represent each neuron with a corresponding qubit yielding a quantum circuit on $M=\sum ^{L+1}_{l=0} m_l$ qubits.
The network's $2^M$-dimensional Hilbert space is the tensor product of $M$ single-qubit Hilbert spaces.
The quantum perceptron $U^l_j$ is composed of two-qubit canonical gates $\can(t_1, t_2, t_3)$ \cite{crooks2019gradients}. To guarantee the network's universality, additional single-qubit gates $u(\theta, \phi, \lambda)$ are applied to all qubits.
The canonical gate is composed of three two-qubit gates:
\begin{equation}\label{can_definition}
\begin{split}
    \can (t_1, t_2, t_3) &= e^{-i\frac{\pi}{2}t_1 X\otimes X} e^{-i\frac{\pi}{2}t_2 Y\otimes Y} e^{-i\frac{\pi}{2}t_3 Z\otimes Z} \\
    &= \rxx (t_1 \pi) \ryy (t_2 \pi) \rzz (t_3 \pi),
\end{split}
\end{equation}
where $X = \big(\begin{smallmatrix}
  0 & 1\\
  1 & 0
\end{smallmatrix}\big),$ 
$Y = \big(\begin{smallmatrix}
  0 & -i\\
  i & 0
\end{smallmatrix}\big),$ and  
$ Z = \big(\begin{smallmatrix}
  1 & 0\\
  0 & -1
\end{smallmatrix}\big)$ are the Pauli matrices and $t_{1,2,3}\in\mathbb{R}$.
The single-qubit gate $u(\theta, \phi, \lambda)$ is defined as
\begin{equation}\label{u_definition}
    u(\theta, \phi, \lambda) =  \begin{pmatrix}
                                    \cos (\theta /2) & -e^{i\lambda}\sin (\theta /2)\\
                                    e^{i \phi} \sin (\theta /2) & e^{i(\lambda+\phi)} \cos (\theta /2)
                                \end{pmatrix},
\end{equation}
where $\theta, \phi, \lambda \in \mathbb{R}$.

The quantum circuit simulating a DQNN is structured layer-wise.
The first $m$ qubits are initialized in a given state $\ket{\phi^\mathrm{in}}$, while all remaining qubits are initialized in $\ket{0}$.
The $u$ gates are  applied first, layer by layer ($l=1,...,L+1$), to all $m_{l-1}$ input qubits. After that, the layer unitary $U^l = \prod ^1_{m_l} U_j^l$ is applied to all input and output qubits. Here, $U_j^l$ is a sequence of $m_{l-1}$ $\can$ gates where the $i$th $\can$ gate acts on the $i$th input and the $j$th output qubit. After each layer $l$, the $m_{l-1}$ input qubits are neglected. The $m_l$ output qubits of this layer serve as the input qubits for the next layer $l+1$. By this, the partial trace of \cref{output_state_definition} is realized. After the output layer $L+1$, again, $u$ gates are applied to the remaining $m$ output qubits.
Thus, the quantum circuit consists of $N_p = 3m + 3\sum ^{L+1}_{l=1} m_{l-1} (1+m_{l})$ parameters.
This procedure is illustrated in Fig.~\hyperref[fig:circuits]{1b} for a \begin{tikzpicture}[yscale=0.7,xscale=0.6, baseline]
		\node(1) [circle,draw,inner sep=0pt,minimum size=4.5pt] at (-1,0) {};
		\node(2) [circle,draw,inner sep=0pt,minimum size=4.5pt] at (-1,0.3) {};
		\node(4) [circle,draw,inner sep=0pt,minimum size=4.5pt] at (0,-0.15) {};
		\node(5) [circle,draw,inner sep=0pt,minimum size=4.5pt] at (0,0.15) {};
		\node(6) [circle,draw,inner sep=0pt,minimum size=4.5pt] at (0,0.45) {};
		\node(7) [circle,draw,inner sep=0pt,minimum size=4.5pt] at (1,0) {};
		\node(8) [circle,draw,inner sep=0pt,minimum size=4.5pt] at (1,0.3) {};
		\draw (1)--(4);
		\draw (2)--(4);
		\draw (1)--(5);
		\draw (2)--(5);
		\draw (1)--(6);
		\draw (2)--(6);
		\draw (4)--(7);
		\draw (5)--(7);
		\draw (6)--(7);
		\draw (4)--(8);
		\draw (5)--(8);
		\draw (6)--(8);
\end{tikzpicture}
DQNN.

\subsection{QAOA}
The QAOA consists of a sequence of alternating unitary operators $e^{-iAt_l}, e^{-iB\tau_l} \in U(d)$, where $t_l,\tau_l \in \mathbb{R}$ and $A$ and $B$ are hermitian matrices initially randomly generated from the gaussian unitary ensemble \cite{farhi2014quantum}. 
The total action of $N$ such operator pairs can be written as 
\begin{equation}
    \mathcal{U} = e^{-iB\tau_N} e^{-iAt_N} \cdots e^{-iB\tau_1} e^{-iAt_1}
\end{equation}
so that the output state is given by $\rho^{\rm out} = \mathcal{U} \ket{\phi^\mathrm{in}}\bra{\phi^\mathrm{in}}\mathcal{U}^\dagger$.

The QAOA can be interpreted as an $\left(N+1\right)$-layer QNN featuring $m$ neurons per layer.
We identify each operator pair $e^{-iAt_l}$, $e^{-iB\tau_l}$ as the layer $l+1$ unitary acting on all $m$ neurons of layer $l$. The resulting state is the input for the neurons of the next layer $l+1$.

\subsubsection*{The Quantum Circuit}
The QAOA's quantum circuit is implemented using only $m$ qubits as each layer is represented by the same set of qubits.
The Hilbert space is the same as for the DQNN but with lower dimension $2^m$.
We initialize the $m$ qubits in a given state $\ket{\phi^\mathrm{in}}$ with dimension $d = 2^m$. $N$ is chosen to be $d^2/2$ so that the QNN converges to an optimal solution \cite{kiani2020learning}. Thus, the number of total parameters equals $N_p = d^2=4^m$.
An exemplary quantum circuit for $m=2$ is depicted in Fig.~\hyperref[fig:circuits]{1c}.

\section{Training the networks}\label{chap3}
In this section we describe the learning algorithm applied and present its implementation.

\subsection{Learning Algorithm}\label{chap3-sec1}

For both QNNs, the task is to learn an unknown unitary $V \in U(2^m)$ from a given training set $\{\ket{\phi^\mathrm{in}_x},\ket{\phi^\mathrm{out}_x} \}^{N_T}_{x=1}$ where $\ket{\phi^\mathrm{out}_x} = V \ket{\phi^\mathrm{in}_x}$.
As a general measure for the network's performance we evaluate the fidelity of the desired output $\ket{\phi ^\mathrm{out}_x}$ and the network's output $\rho ^\mathrm{out}_x$. Averaging over all pairs $x$ in the training set yields the training cost of the network:
\begin{equation}\label{cost_definition}
	C_T=\frac{1}{N_T}\sum_{x=1}^{N_T} \langle\phi^{\text{out}}_x\rvert\rho_x^{\text{out}}\lvert\phi^{\text{out}}_x\rangle.
\end{equation}

The networks are trained iteratively by gradient descent.
Both are implemented as quantum circuits parameterized by $N_p$ independent parameters. The parameters are initialized as $\vec{p}_0 = (p_1, ..., p_{N_p})^T$.
At each epoch, all parameters are updated so that they (locally and infinitesimally) maximize the training cost.
The new parameters are given by $\vec{p}_{t+1} = \vec{p}_{t} + \vec{dp}$. The update step is defined as $\vec{dp} = \eta \vec{\nabla} C \left(\vec{p}_t\right)$ with the learning rate $\eta$. The gradient of the training cost is approximated by:
\begin{equation}
    \nabla _k C \left(\vec{p}_t\right) = \frac{C\left(\vec{p}_t + \epsilon\vec{e}_k\right) - C\left(\vec{p}_t - \epsilon\vec{e}_k\right)}{2\epsilon} + \mathcal{O}\left(\epsilon^2\right)
\end{equation}
where $\vec{e}_{k}$ has components $e_{k}^j=\delta_{k}^j$, $k,j=1,...,N_p$ and $\epsilon > 0$.
In the following, $\epsilon$ and $\eta$ are called the hyperparameters of our networks.

After training we evaluate the networks' generalization via a validation set $\{\ket{\phi^\mathrm{in}_x},\ket{\phi^\mathrm{out}_x} \}^{N_V}_{x=1}$ built in the same way as the training set. Analogously, the validation cost is defined as:
\begin{equation}\label{validation_cost_definition}
	C_V=\frac{1}{N_V}\sum_{x=1}^{N_V} \langle\phi^{\text{out}}_x\rvert\rho_x^{\text{out}}\lvert\phi^{\text{out}}_x\rangle.
\end{equation}

\subsection{Implementation}\label{implementation}

The execution of the QNNs involves three crucial steps.

\textit{i. Initialization}: First, a set of training pairs as well as a set of validation pairs, if not already given, are generated from a target unitary $V\in U\left(d\right)$. The hyperparameters are set according to the problem and network at hand. The parameters $\vec{p}_0$ are initialized randomly.

\textit{ii. Learning}: The following steps are repeated for a given number of epochs. The full quantum circuit (see Fig.~\hyperref[fig:circuits]{1a}) consists of three main parts.
In the first part, a set of $m$ qubits is initialized in the state $\ket{\phi^\mathrm{out}_x}$ and another $m$ qubits in $\ket{\phi^\mathrm{in}_x}$.
The remaining qubits, if necessary, are initialized in $\ket{0}$. The second part features the network's evaluation of $\ket{\phi^\mathrm{in}_x}$. In the last part, the fidelity is calculated using the destructive swap test \cite{cincio2018learning, garcia2013swap}. The full circuit is executed for all states in the training set to calculate the training cost. The network parameters are updated using gradient descent as described in \cref{chap3-sec1}.

\textit{iii. Validation}:
The network is evaluated using a validation set. Its states are unknown to the network and uncorrelated to the training set.
The resulting validation cost yields a measure for the networks' generalization capabilities.

\section{Results}\label{chap4}
In this section, we perform numerical studies of the previously described networks.
This includes an analysis of their generalization capabilities and noise robustness.
Additionally, we train both networks on real quantum computers. 

\subsection{Setup} \label{chap4-part1}
To analyze and compare the two networks described in \Cref{chap2}, we choose a target unitary of dimension $d=4$.
Thus, both networks have $m=2$ input and output qubits, respectively.
We train a one-layer DQNN of the form
\begin{tikzpicture}[yscale=0.7,xscale=0.6, baseline]
\node(1) [circle,draw,inner sep=0pt,minimum size=4.5pt] at (-1,0) {};
\node(2) [circle,draw,inner sep=0pt,minimum size=4.5pt] at (-1,0.3) {};
\node(4) [circle,draw,inner sep=0pt,minimum size=4.5pt] at (0,0) {};
\node(5) [circle,draw,inner sep=0pt,minimum size=4.5pt] at (0,0.3) {};
\draw (1)--(4);
\draw (2)--(4);
\draw (1)--(5);
\draw (2)--(5);
\end{tikzpicture}
and a QAOA network with $N=8$ layers.
These shallow network architectures are well suited for an analysis of both network types. They already exhibit the quantum entanglement and implement the basic structure of QNNs. This enables a generalization of the main results to more extensive networks.
These networks are implemented using 6 (4) qubits for the DQNN (QAOA), including the initialization and measurement as described in \Cref{implementation}. Training these networks requires the learning of 24 (16) parameters, respectively.
Due to the small number of parameters for both networks, we do not encounter the effects of barren plateaus.
For the DQNN, the parameters are initialized in the range $\left[0,2\pi\right)$. Here, we found $\eta = 0.5$ and $\epsilon = 0.25$ as the optimal hyperparameters.
For the QAOA, we initialize the parameters in the range $\left[-1,1\right]$ and found the hyperparameters $\eta = 0.075$ and $\epsilon = 0.05$ optimal.

The implementation and execution of the corresponding quantum circuits was carried out using the open-source SDK Qiskit \cite{Qiskit} and the quantum devices hosted by IBM \cite{ibmqe}.
Qiskit allows decomposition of the parameterized networks mentioned in \Cref{chap2} to a predefined set of elementary gates yielding a quantum circuit that the IBM quantum devices execute. 
While the number of required basis gates for the initialization (6-12 single-qubit gates and 6 CNOT gates) and the fidelity measurement (2 single-qubit gates and 2 CNOT gates) is independent of the network type, the required qubits for implementing the core network circuit differs.
For example, mapping the network's circuit to the H-topology of ibmq\_casablanca \cite{ibmqe} results in 57 (97) single-qubit gates and 63 (57) CNOT gates for the DQNN (QAOA).

The training of the network was executed in a hybrid manner. At each epoch, the cost was evaluated by the quantum execution, which was then used to update the parameters classically. 
To guarantee the training's success, it was necessary to use a training set that was large enough to determine the transformation of the target unitary. For a 2-2 network ($V\in U\left(4\right)$), this was achieved using four training pairs \cite{poland2020no}.

\begin{figure*}[ht]
\centering
\rule{\linewidth}{0pt}
\input{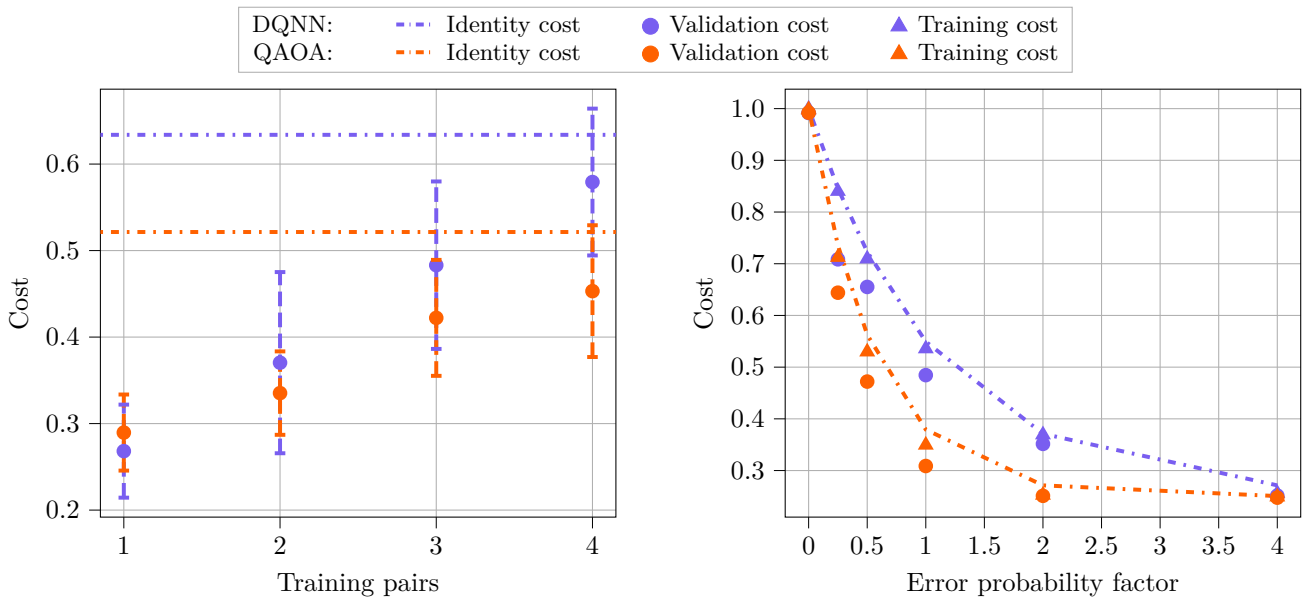}
\caption{\textbf{Training on simulated devices.} (a) shows the validation cost (\protect\validationCostMarkOnly) after training both networks with $N_T=1,\dots,4$ training pairs for 1000 epochs. The identity cost (\protect\calibrationline) gives an estimate for the cost of an ideally trained network. Although the networks use identical device simulations, the DQNN generally reaches higher validation costs. In (b), the networks are trained for varying gate noise until the validation cost convergences, whereby the training cost (\protect\trainingCostMarkOnly) is shown, additionally. Here, $k=1$ corresponds to the gate noise of currently available NISQ devices. Especially in this region, the DQNN performs better.
In both analyses, the quantities are averaged over 20 (a) and 5 (b) training sessions with different target unitaries $V$ and start parameters $\vec{p_0}$.}
\label{simulateddevice}
\end{figure*}
\subsection{Generalization Analysis}\label{chap4-part2}

Due to the generally limited amount of quantum data, it is of paramount interest to analyze the network's generalization capabilities, i.e., how does it perform on previously unseen data. We quantify this via a \emph{validation cost} defined by four validation pairs while training the network using only $N_T =1, \dots, 4$ training pairs. The validation cost represents a fair metric to compare QNN architectures as it quantifies the influence of \emph{overfitting} where a QNN architecture maximises the training cost at the expense of performance on unseen data.

We executed the quantum circuits on a simulator incorporating the real-time noise of \textit{ibmq\_casablanca}, which we set constant for all values of $N_T$. 
The training was repeated several times to average over different initial parameters, target unitaries, and noise snapshots. 

The results of this study are shown in \cref{fig:generalization}.
Alongside the validation costs as a function of the number of training pairs, the networks' \emph{identity costs} are also shown. The identity cost is calculated using $V=\mathbbm{1}$ and corresponding parameters so that the respective network acts as the identity.
Each gate is still applied and adds noise to the circuit.
Thus, the identity cost provides an estimate for the best possible costs of an ideally trained network.

Our studies show that both networks are capable of generalizing the available information.
As expected, the validation costs increase with the number of training pairs.
According to the higher identity cost, the DQNN also reaches higher validation costs than the QAOA.
Additionally, the DQNN generalizes with higher reliability as its different training sessions for $N_T=4$ yield very few outliers.
In contrast, the QAOA's validation cost is approximately uniformly distributed around the mean.

\subsection{Gate Noise Analysis}\label{chap4-part3}
The success of the DQNN and the QAOA is strongly affected by the noise of the executing quantum device.
The primary sources are readout noise and gate noise \cite{nachman2020unfolding}.
Both networks are equally influenced by readout noise as they are evaluated by the same measurement method.
Our focus in this paper is on the comparison of both networks and thus we neglect readout noise.
The influence of gate noise, however, does differ because the networks consist of different gates.
To test the influence of gate noise on both networks' performances, we trained the DQNN and the QAOA
using a depolarizing quantum error channel \cite{nielsen2002quantum}. 
It was parametrized by the depolarization probabilities
$\lambda^\text{g}=k \lambda^\text{g}_0$ with basis gates $g=\text{CNOT, SX, RZ}$ and scaling factor $k$. The parameter
$\lambda^\text{g}_0$ was chosen to match the gate error probabilities of
current NISQ devices.
Here, we chose $\lambda^\text{CNOT}_0 =\num[scientific-notation=true]{3.14e-2}$, $\lambda^\text{SX}_0 =\num[scientific-notation=true]{1.18e-3}$, $\lambda^\text{RZ}_0=0$, which is an approximation for \textit{ibmq\_16\_melbourne} \cite{ibmqe}.
We also used \textit{ibmq\_16\_melbourne's} qubit coupling map.

\Cref{fig:errorrobustness} shows the numerical results comparing the DQNN and the QAOA for different error probability factors $k$. Note that $k=1$ corresponds to the gate noise of currently available NISQ devices.
For $k=0$, i.e., a noise-free quantum device, both QNNs perform equally well.
In the presence of gate noise, however, the DQNN has a higher identity cost, which in each case also yields a higher training and validation cost.
This indicates that the DQNN is more suitable for learning unitaries on NISQ devices than the QAOA.

\begin{figure*}[ht]
\centering
\rule{\linewidth}{0pt}
\input{realdevice_figure.tex}
\caption{\textbf{Training on quantum computers.} This figure shows the relevant quantities measured while training the DQNN (a) and the QAOA (b).
Both networks have been trained for $100$ epochs. Each epoch, the training cost (\protect\trainingCostMark) is calculated via four $4$ training pairs. Every fifth epoch, the validation cost (\protect\validationCostMark) is measured using $4$ validation pairs, as well as the identity cost (\protect \calibrationCostMark) using $4$ output training states.
}
\label{fig:realdevice}
\end{figure*}
\subsection{NISQ Device Execution} \label{chap4-part4}
Apart from simulating the QNNs to study their main properties, we have to execute them on actual NISQ devices to broaden our knowledge about their real-world applications and benefit from their advantages.
To demonstrate both the challenges and the potential of training QNNs on NISQ devices, we present results of a single training session, respectively. 
By this, we can evaluate the direct effect of the real device noise on the different costs.
We trained both QNNs using the IBM 7-qubit device \textit{ibmq\_casablanca} \cite{ibmqe}.

The relevant costs during training the DQNN and the QAOA to learn an unknown unitary $V\in U(d)$ are shown in \cref{fig:realdevice}.
In the noise-free case, the training cost should always be monotone increasing for well-chosen hyperparameters. Here, the variance in the training cost indicates the networks' response to gate and measurement noise.
However, the training and validation costs seem to be correlated to the varying identity cost. That indicates that the variance of the identity cost is not only a statistical error but arises from a noise drift of the quantum device itself.
In the previous analyses, the identity cost served as the upper limit for the training and validation cost. Here, the training cost exceeds the identity cost after a few epochs.
This indicates the networks' surprising ability to factor in the noise circumstances of the real device.
Having noted this, the high validation to identity cost ratio demonstrates the remarkable capability of both networks to generalize the provided information despite the high noise levels.


\section{Discussion and Outlook}\label{chap5}
In this paper we introduced a \textit{dissipative} Quantum Neural Network ansatz and compared it to the Quantum Approximate Optimization Algorithm.
Both QNNs were implemented and trained on IBM's quantum computers.

It was found that a DQNN learns an unknown unitary better than the QAOA, as quantified by the validation cost.
Our analysis demonstrates that the DQNN generalizes with higher reliability and accuracy than the QAOA.
Moreover, we found that the DQNN is less susceptible to gate noise than the QAOA, most notably in the region of current NISQ device noise levels.
We demonstrated the ability of both networks and the gradient descent-based learning algorithm to withstand the noise of today's quantum devices. However, it should be noted that noise still prevents QNNs from reaching high fidelities.

Besides the reduction of noise, the further development of quantum devices holds potential in providing resettable qubits. This would reduce the number of qubits required to operate a DQNN and thus allow the exploration of multiple layers.
It is a fascinating problem whether DQNN for higher-dimensional unitaries and non-unitary maps offer similar advantages. We leave this question to future work.

The code is available at \url{https://github.com/qigitphannover/DeepQuantumNeuralNetworks}.

\vspace{5pt}
\begin{acknowledgments}
Helpful discussions with Dmytro Bondarenko, Terry Farrelly, Polina Feldmann, Alexander Hahn, Jan Hendrik Pfau, Robert Salzmann, Daniel Scheiermann, Viktoria Schmiesing, Marvin Schwiering, and Ramona Wolf are gratefully acknowledged. This work was supported, in part, by the Quantum Valley Lower Saxony (QVLS), the DFG through SFB 1227 (DQ-mat), the RTG 1991, and funded by the Deutsche Forschungsgemeinschaft (DFG, German Research Foundation) under Germany's Excellence Strategy EXC-2123 QuantumFrontiers 390837967.
We acknowledge the use of IBM Quantum services for this work. The views expressed are those of the authors, and do not reflect the official policy or position of IBM or the IBM Quantum team.
\end{acknowledgments}


\bibliographystyle{apsrev4-2}
\bibliography{BibFile}
\end{document}